\documentclass{article}

\usepackage{arxiv}
\usepackage[utf8]{inputenc} 
\usepackage[T1]{fontenc}    
\usepackage{amsmath}
\usepackage{hyperref}       
\usepackage{url}            
\usepackage{booktabs}       
\usepackage{amsfonts}       
\usepackage{nicefrac}       
\usepackage{microtype}      
\usepackage{lipsum}
\usepackage{graphicx}
\graphicspath{ {./images/} }
\usepackage{amsthm}

\usepackage{todonotes}

\usepackage{listings}
\usepackage{color}

\definecolor{dkgreen}{rgb}{0,0.6,0}
\definecolor{gray}{rgb}{0.5,0.5,0.5}
\definecolor{mauve}{rgb}{0.58,0,0.82}

\title{Effects of position-dependent mass (PDM) on the bound-state solutions of a massive spin-0 particle subjected to the Yukawa potential}

\author{
	P. H. F. Oliveira \\
	Instituto de Desenvolvimento Rural,\\ Universidade da Integração Internacional da Lusofonia Afro-Brasileira\\
	Campus das Auroras, 62790-000, Redenção, Ceará, Brazil \\
	\href{mailto:pedrooliveira@fisica.ufc.br}{pedrooliveira@fisica.ufc.br} \\
	\And
	W. P. Lima \\
	Departamento de Física, Universidade Federal do Ceará\\
	Campus do Pici, 60455-760, Fortaleza, Ceará, Brazil \\
	\href{mailto:wellisson@fisica.ufc.br}{wellisson@fisica.ufc.br} \\
}

\begin{document}
\maketitle
\begin{abstract}
With the advent of Albert Einstein's theory of special relativity, Klein and Gordon made the first attempt to elevate time to the status of a coordinate in the Schrödinger equation. In this study, we graphically discuss the eigenfunctions and eigenenergies of the Klein-Gordon equation with a Yukawa-type potential (YP), within a position-dependent mass (PDM) framework. We conclude that the PDM leads to the equivalence of the positive ($E^+$) and negative ($E^-$) solution states at low energies. We observe that in the energy spectrum as a function of $\eta$ (YP intensity factor), the PDM can induce gap closure at the critical point where $E^+$ and $E^-$ become imaginary. In the spectrum as a function of $\alpha$ (YP shielding factor), it can compel the energies to be zero at $\alpha=0$, instead of being equal to $(m_0c^2)$ as in the invariant mass case.
\end{abstract}

\newpage

\section{Introduction}\label{sec:intro}

In quantum theory, specifically in quantum field theory, profound connections are established between fields and particles with a given spin through the spin-statistics theorem \cite{zee2010quantum}. This theorem states that particles with integer spin are described by a commutation algebra, while particles with half-integer spin obey an anticommutation algebra \cite{netoteoria}. In particular, spin-zero particles such as the mesons $\pi^0$ are described by real scalar fields, while mesons $\pi^{\pm}$ are described by complex scalar fields \cite{rubakov2009classical}.

The dynamics of the real scalar field are governed by the Klein-Gordon equation, in a manifestly relativistic formulation \cite{desai2010quantum}
\begin{equation}
	p^\mu p_\mu\psi=m_0^2c^2\psi,
\end{equation}
where $m_0$ is the rest mass of the Klein-Gordon particle.

Using Dirac's prescription in quantum mechanics and considering a minimal coupling with the electromagnetic field \cite{rubakov2009classical}, the stationary radial solution is obtained as the solution to the equation \cite{bromley2012quantum}
\begin{equation}
	\left[\frac{d^2}{dr^2}-\frac{l(l+1)}{r^2}+k^2\right]\phi(r)=0,
\end{equation}
where
\begin{equation}
	k^2=\frac{\left[E-V(r)\right]^2-m_0^2c^4}{\hbar^2c^2}.
\end{equation}

Obtaining the exact solution of the Klein-Gordon equation for any potential is relevant in theoretical physics. However, receiving the solution analytically has only been possible for certain specific interaction potentials. The literature generally employs the Niforov-Uvarov method \cite{Berkdemir12}, which requires that the scalar and vector components of the potential have the same functional form \cite{ALHAIDARI200687}. Some potentials that have been solved in this manner include: Hulthen \cite{Farrokh:2013jsa}, Woods-Saxon \cite{doi:10.1142/S0129183108012480}, Rosen-Morse \cite{akbariyeh2008exact}, and Yukawa \cite{ntibi2020relativistic}.

The Yukawa potential, like other short-range potentials, can be used to model nuclear interactions \cite{Yukawa:1935xg}, electrostatics in colloidal media \cite{RAMOS2018}, describe the Meissner effect in superconductors \cite{jackson2012classical}, and in massive field theories \cite{phbfmodel}. Analytically, the Yukawa potential can be obtained as the spherically symmetric solution of a stationary point charge in the Maxwell-Proca equation \cite{Proca}
\begin{equation}
	(\nabla^2-\alpha^2)V(r)=\eta\delta(r)\Rightarrow V(r)=-\eta\frac{e^{-\alpha r}}{r},
\end{equation}
the Yukawa potential is thus the shielded Coulomb potential, with its strength diminishing as $\alpha$ increases.

In this article, we will discuss the bound state solutions of the Klein-Gordon equation with Yukawa interaction, without resorting to the traditionally employed method. This solution was obtained accurately in Ref. \cite{Arda:2011zz}, utilizing the PDM formalism; however, it did not analyze the various configurations or extract interpretations of the PDM effects on the behavior of the wave functions and energy spectra.


\section{Bound State Solutions}

A relevant approximation in the literature, to avoid the use of integrals in complex variables when working with potentials of the form $V(r)\propto r^{-t}$, initially in the Dirac equation \cite{ALHAIDARI200472} and later in the Klein-Gordon equation, is the approximation between the Coulomb and Hulthen potentials for $\alpha\ll1$ \cite{PhysRevA.14.2363}
\begin{equation}
    \lim\limits_{\alpha\rightarrow0}\left[4\alpha^2\frac{e^{-2\alpha}}{(1-e^{-2\alpha r})^2}\right]=\frac{1}{r^2}.
\end{equation}

With this approximation, it is possible to rewrite the Yukawa potential in the form
\begin{equation} 
    V(r)=-2\alpha\eta\frac{e^{-2\alpha r}}{1-e^{-2\alpha r}}.
\end{equation}

\begin{figure}
    \centering
	{\includegraphics[scale=0.8]{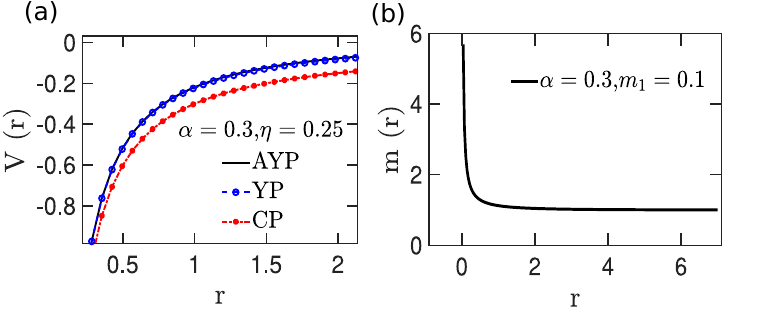}}
	\caption{(a) Comparison between the rewritten Yukawa potential (solid black line), the Yukawa potential (blue dashed line with circles), and the Coulomb potential (red dot-dashed line with asterisks), all versus position, with $\alpha=0.3$ and $\eta=0.25$. (b) Mass function versus position, with $\alpha=0.3$ and $m_1=0.1$.}
	\label{espectro_n}
\end{figure}

Indeed, for small values of $\alpha$, such as $0.3$, there is excellent agreement with the exact Yukawa potential. This fact is illustrated, along with the Coulomb potential, in Fig. \ref{espectro_n}-a.

On the other hand, the PDM (Position-Dependent Mass) formalism has many applications in different areas, such as impurities in crystals \cite{slater1949electrons}, quantum wells and dots \cite{serra1997spin}, and semiconductor heterostructures \cite{gora1969theory}, and has been considered in the study of some potentials \cite{jia2007trapping,jia2008extension,alhaidari2004solution}. Therefore, the mass function is defined as
\begin{equation}
    m_0\rightarrow m(r)=m_0+\frac{m_1}{e^{2\alpha r}-1}, 
\end{equation}
which exhibits a response effect to the potential, illustrated in Fig. \ref{espectro_n}-b.

In Ref. \cite{Arda:2011zz}, a variable change of the form $r\rightarrow z=(1-e^{-2\alpha r})^{-1}$ was employed, along with the assumption of a solution of the type

\begin{equation} \phi(z)=z^{\lambda_1}(1-z)^{\lambda_2}\psi(z). \end{equation}

This approach transforms the Klein-Gordon equation into the following form
\begin{eqnarray}\label{eqkgpdm}
    z(1-z)\psi''(z) &+& [1+2\lambda_1-2(\lambda_1+\lambda_2+1)z]\psi'(z)\nonumber\\&&+\left[-\lambda_1^2-\lambda_2^2-\lambda_1-\lambda_2-2\lambda_1\lambda_2+l(l+1)+\frac{\beta^2m_1^2c^4}{4\alpha^2}-\beta^2\eta^2\right]\psi(z)=0, 
\end{eqnarray}
where the parameters are given by
\begin{subequations}\label{param}
    \begin{eqnarray}
        \lambda_1^2 &=& \frac{\beta^2}{4\alpha^2}(m_0^2c^4-E^2)+\frac{E\beta^2\eta}{\alpha}+\Lambda(\beta),\label{eq_lambda1}\\
        \lambda_2^2 &=& \frac{\beta^2}{4\alpha^2}(m_0^2c^4-E^2),\label{param2}\\
        \Lambda(\beta) &=& -\frac{\beta^2m_0m_1c^4}{2\alpha^2}+\frac{\beta^2m_1^2c^4}{4\alpha^2}-\beta^2\eta^2.
    \end{eqnarray}
\end{subequations}

The solution of the differential equation \eqref{eqkgpdm} is the Gauss hypergeometric function ${}_2F_1(\xi_1,\xi_2;\xi_3;z)$ and, therefore, since its DE can be mapped into a Sturm-Liouville problem \cite{Everitt2005}, it is guaranteed that: (i) the energies of the particle are real; (ii) the eigenfunctions of the Hamiltonian are orthonormal; and (iii) the eigenstates form a basis in Hilbert space \cite{philipetesepaper}.

Indeed, the parameters $\xi$ are determined by
\begin{subequations}
\begin{eqnarray}
    \xi_1 &=& \lambda_1+\lambda_2+\frac{1}{2}[1+\mathcal{L}(l)],\\
    \xi_2 &=& \lambda_1+\lambda_2+\frac{1}{2}[1-\mathcal{L}(l)],\\ 
    \mathcal{L}(l) &=& \sqrt{1+4l(l+1)+\frac{\beta^2m_1^2c^4}{\alpha^2}-4\beta^2\eta^2}. 
\end{eqnarray} \end{subequations}

For $\phi$ to represent the wave function of a particle, it is necessary to ensure that it is square-integrable \cite{sakurai2017modern}. Therefore, restrictions must be imposed on the parameters. A mathematical condition in ${}_2F_1(a,b;c;x)$ ensures that if $a$ or $b$ (since the function is symmetric in the permutation $a\leftrightarrow b$) is a negative number, the hypergeometric function reaches a finite value \cite{bagdasaryan2009note}. Thus, the quantum condition is given by $\xi_1=-n$ ($n=0,1,\cdots$) \cite{Arda:2011zz}. Therefore, the radial solution of the Klein-Gordon equation subjected to the Yukawa potential via the PDM formalism, in the variable $z\rightarrow z=(1-s)/2$, is given by
\begin{equation}
    \phi\left(\frac{1-s}{2}\right)=N'(1-s)^{\lambda_1}(1+s)^{\lambda_2} {}_2F_1\left(-n,n+2\lambda_1+2\lambda_2+1;1+2\lambda_1;\frac{1-s}{2}\right),
\end{equation}
where $N'=2^{-(\lambda_1+\lambda_2)}N$ is the normalization constant given by
\begin{equation} |N'|^2=\frac{2\lambda_1+2k+1}{|\Sigma(n,k)|^2{}_2F_1(-2\lambda_2,1;2\lambda_1+2k+2;-1)}, \end{equation}
and where
\begin{equation}
    \Sigma(n,k)\equiv\Gamma(1+2\lambda_1)\sum_{k=0}^{n}\frac{\Gamma(-n+k)\Gamma(2\lambda_1+2\lambda_2+n+k+1)2^{-k}}{\Gamma(-n)\Gamma(2\lambda_1+2\lambda_2+n+1)\Gamma(2\lambda_1+k+1)k!}.
\end{equation}

The Table \ref{tab:tableN} displays the values of $N'$ for the $s$ states ($l=0$) of ``positive'' and ``negative'' energies in the cases $m_1 = 0$ and $m_1 = 0.1$ for a visualization of the effect of PDM on the normalization constant of the wave functions.

\begin{table}[!h]
    \centering
    \begin{tabular}{|c|c|c|c|c|}
        \hline
        \textbf{$n$} & \multicolumn{2}{c|}{$N' [\phi_{n,0}^+]$} & \multicolumn{2}{c|}{$N' [\phi_{n,0}^-]$} \\ 
        \hline
        & \textbf{$m_1 = 0$} & \textbf{$m_1 = 0.1$} & \textbf{$m_1 = 0$} & \textbf{$m_1 = 0.1$} \\ 
        \hline 
        $0$ & $2.02568$ & $2.14417$ & $1.25551$ & $2.15323$ \\ 
        \hline
        $1$ & $4.78012$ & $23.2147$ & $0.89947$ & $23.5816$ \\ 
        \hline
        $2$ & $15.1767$ & $125.661$ & $2.76688$ & $129.296$ \\ 
        \hline
        $3$ & $42.3247$ & $425.014$ & $11.1521$ & $444.511$ \\ 
        \hline
        $4$ & $112.618$ & $1028.11$ & $34.4452$ & $1097.33$ \\ 
        \hline
        $5$ & $271.253$ & $1822.97$ & $105.462$ & $1991.1$ \\ 
        \hline
        $6$ & $704.911$ & $2345.5$ & $288.54$ & $2639.1$ \\ 
        \hline
        $7$ & $1721.39$ & $2050.48$ & $821.447$ & $4327.26$ \\ 
        \hline
        $8$ & $4501.8$ & $136.028$ & $2196.82$ & \textit{Im. Puro} \\ 
        \hline
        $9$ & $11130.9$ & \textit{Im. Puro} & $6001.23$ & \textit{Im. Puro} \\ 
        \hline
        $10$ & $21574.6$ & \textit{Im. Puro} & $13918.7$ & \textit{Im. Puro} \\ 
        \hline
    \end{tabular}
    \caption{The values of $N'$ that normalize the states $s$ ${\phi^\pm_{n,0}}$ for different values of $m_1$ were determined, considering the parameters $\alpha = 0.01$, $\eta = 0.1$, and $m_0 = 1$. Source: The authors.}
    \label{tab:tableN}
\end{table}

\begin{figure}
    \centering
    \includegraphics[width=0.5\linewidth]{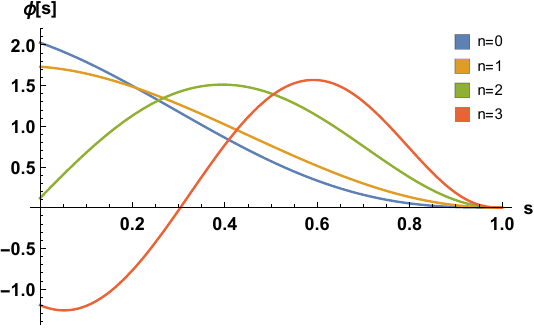}
    \label{fig:sub1}
    \caption{Positive solutions of the `$s$' states ($l=0$), with $n=\{0,1,2,3\}$. Source: The authors.}
    \label{fig:phis1e2pos}
\end{figure}

Fig. \ref{fig:phis1e2pos} illustrates the behavior of the positive energy wave functions for the first states. By neglecting the centrifugal effects, it is observed that the tendency to become less localized at the origin, as $n$ increases, is reflected in $|\phi|^2$. Although $|\phi|^2$ cannot be strictly interpreted as a probability density, it is related to how the charge is distributed spatially \cite{bromley2012quantum}.

\section{Discussion of the results}

\subsection{Effects of the PDM formalism on the eigenfunctions}

\begin{figure}[!h]
    \centering
    \includegraphics[width=0.5\linewidth]{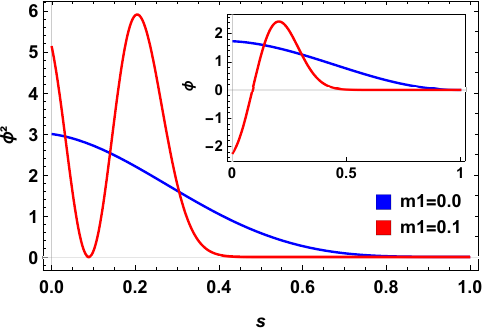}
    \caption{Effects of $m_1$ on positive energy solutions, $n=1$, $l=0$, $\eta=0.1$, and $\alpha=0.01$. Source: The authors.}
    \label{fig:n1pos-eps-converted-to}
\end{figure}

\begin{figure}[!h]
    \centering
    \includegraphics[width=0.5\linewidth]{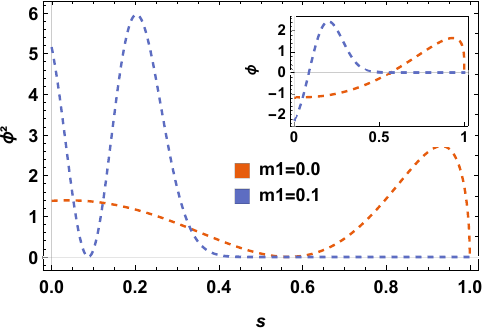}
    \caption{Effects of $m_1$ on negative energy solutions, $n=1$, $l=0$, $\eta=0.1$, and $\alpha=0.01$. Source: The authors.}
    \label{fig:n1neg-eps-converted-to}
\end{figure}

In addition to providing a method for analytically solving the Klein-Gordon equation for the Yukawa potential, as shown by \cite{Arda:2011zz}, the effect of PDM manifests as a kind of response to the potential. Figs. \ref{fig:n1pos-eps-converted-to} and \ref{fig:n1neg-eps-converted-to} illustrate the behavior of $|\phi(s)|^2$ and highlight the impact of $m_1$ on the ``probability''.

\begin{figure}[!h]
    \centering
    \includegraphics[width=0.5\linewidth]{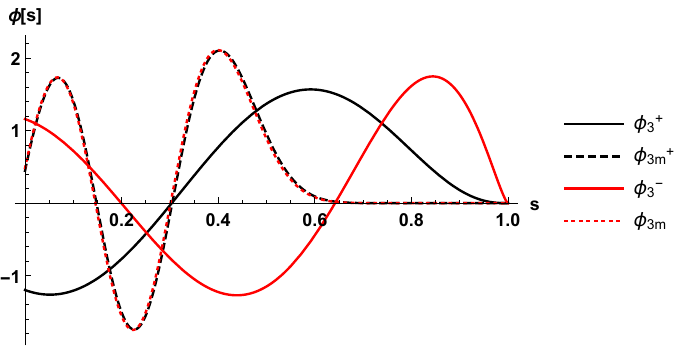}
    \caption{Plot of positive (black) and negative (red) energy solutions, for $m_1=0$ (solid line) and $m_1=0.1$ (dashed line). Source: The authors.}
    \label{fig:n3-eps-converted-to}
\end{figure}

The effect of mass becomes even more intriguing when comparing the pairs of states $\phi^+{n,0}$ and $\phi^-{n,0}$, which generally do not appear to be related. However, the existence of $m_1 \neq 0$ causes them to coincide, as shown in Fig. \ref{fig:n3-eps-converted-to}. This agreement persists up to the state $n=6$.

\subsection{Effects of the PDM formalism on the eigenenergies}

The energy states of the particle as a function of the numbers $(n,l)$ can be determined. It is important to note that, since this is a solution of the Klein-Gordon equation, two energy configurations must be obtained, which will be denoted as $E^+$ and $E^-$, corresponding to the particle and the antiparticle, respectively.

By combining Eqs. \eqref{eq_lambda1} and \eqref{param2}, along with the equality
\begin{equation}
    \lambda_1 + \lambda_2 = -\frac{1}{2}\left[2n + \mathcal{L}(l) + 1\right],
\end{equation}
one can obtain the energies, up to second-order terms in $\alpha$ and $\eta$, given by
\begin{eqnarray}\label{eq_energy}
    E^{(\mp)}_{nl} &=& \frac{\alpha\eta}{2[(\mathcal{L}(l)+2n+1)^2+\beta^2\eta^2]}\Bigg[[\mathcal{L}(l)+2n+1]^2-4\Lambda(\beta)\nonumber\\ &&\mp \frac{\mathcal{L}(l)+2n+1}{\beta\eta}\sqrt{\frac{4\beta^2m_0^2c^4}{\alpha^2}[(\mathcal{L}(l)+2n+1)^2+\beta^2\eta^2]-[(\mathcal{L}(l)+2n+1)^2-4\Lambda(\beta)]^2}\Bigg].
	\end{eqnarray}

Equation \eqref{eq_energy} provides a good approximation for $\eta \leq 0.25$ and $\alpha \leq 0.3$ \cite{arda2011effective}. It can also be compared with the expression for the energy eigenvalues of the corresponding Schrödinger equation \cite{onate2016eigensolutions}, given by
\begin{equation}
    E_{nl} = \frac{\alpha^2\hbar^2l(l+1)}{2m_0} - \frac{\alpha^2\hbar^2}{2m_0}\left[\frac{\frac{-2m_0\eta}{\alpha\hbar^2}-(n+l+1)^2-l(l+1)}{2(l+n+1)}\right]^2.
\end{equation}

\begin{figure}[!h]
    \centering
    \includegraphics[width=0.6\linewidth]{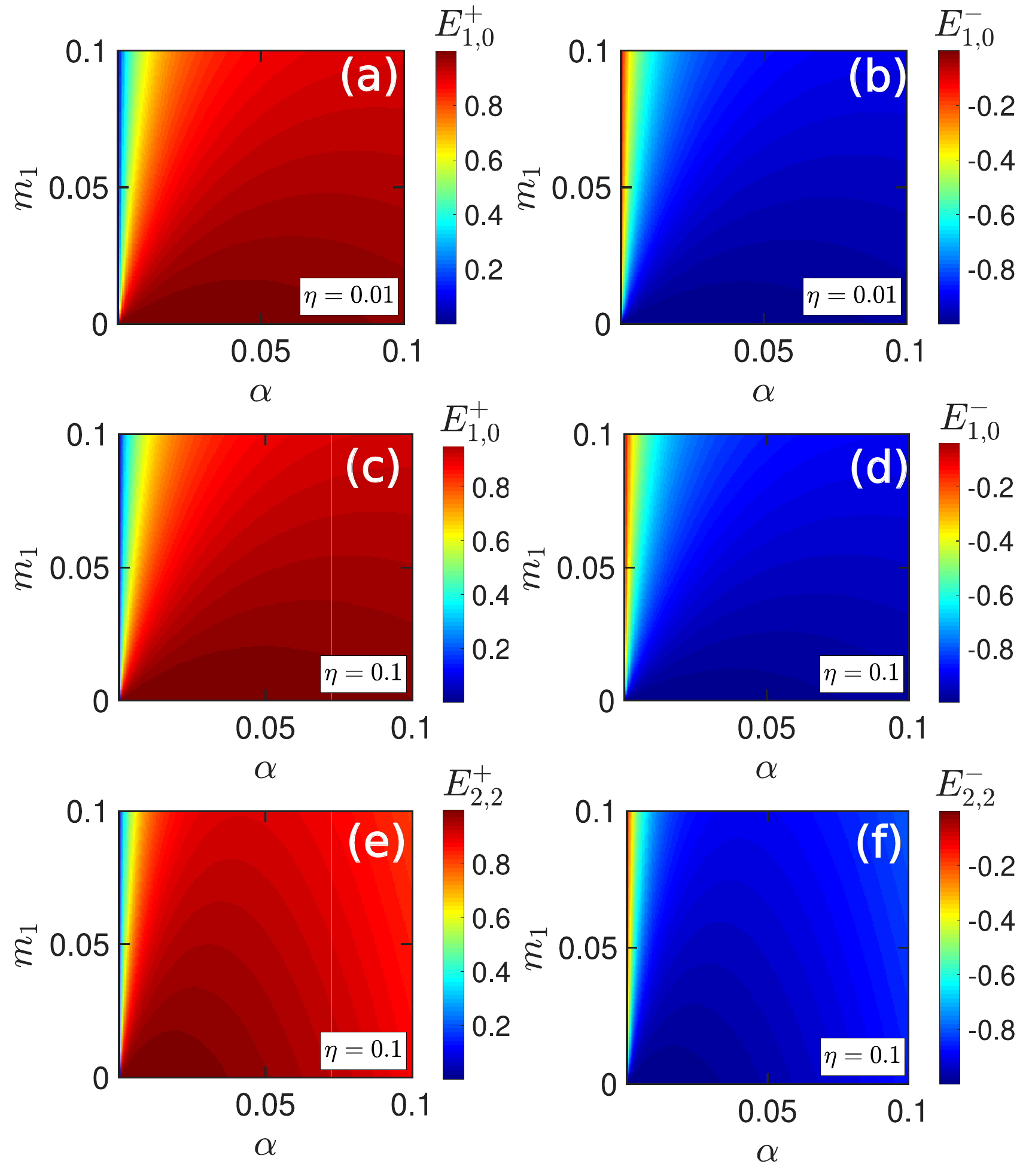}
    \caption{Positive energy level curves on the left and negative on the right, versus $m_1$ versus $\alpha$. In (a-b) we consider $n=1$, $l=0$, and $\eta=0.01$. In (c-d) we also use $n=1$, $l=0$, but set $\eta=0.1$. Finally, in (e-f), we again set $\eta=0.1$, for the case where $n=2$ and $l=2$. Source: The authors.}
    \label{fig:niveis_E_versus_m1_alpha_3D}
\end{figure}

In Fig. \ref{fig:niveis_E_versus_m1_alpha_3D}, we present the energy level curves, with positive energies shown on the left and negative energies on the right, as a function of $m_1$ and $\alpha$. In graphs (a-b), we consider the parameters $n=1$, $l=0$, and $\eta=0.01$. In graphs (c-d), we maintain $n=1$ and $l=0$ while imposing $\eta=0.1$. In graphs (e-f), we again adopt $\eta=0.1$, considering the case where $n=2$ and $l=2$. Immediately, we observe that increasing $\alpha$ (keeping $m_1$ fixed) results in an increase in the magnitudes of $E^+$ and $E^-$ up to a certain point, referred to as the inversion point, where we note an opposite behavior, leading to a decrease in the magnitudes of these energies. The inversion points occur for progressively smaller values of $\alpha$ as $m_1$ decreases and $n$ and $l$ increase. On the other hand, increasing the value of $m_1$ (keeping $\alpha$ fixed) results in a decrease in the magnitudes of the eigenvalues $E^+$ and $E^-$, without an inversion point in this behavior. Furthermore, we can note that as the energy level increases, the energy values corresponding to each configuration point $(m_1,\alpha)$ also increase. Moreover, when considering the same energy level $(n=1,l=1)$, we observe that as the value of $\eta$ increases, the eigenvalue $E^+$ rises while $E^-$ decreases, both in magnitude, resulting in increasingly positive energies.

\begin{figure}[!h]
    \centering
    \includegraphics[width=0.6\linewidth]{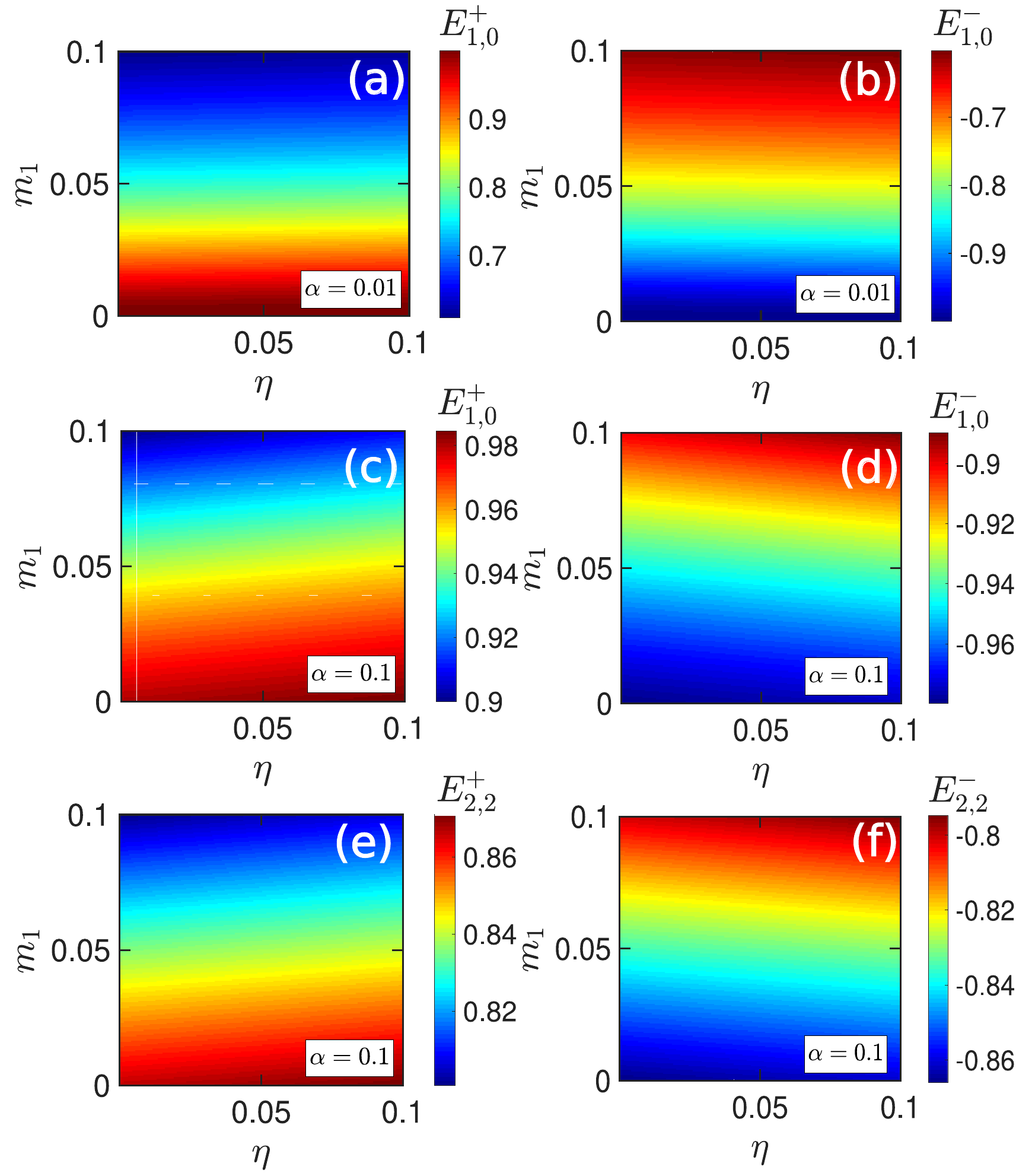}
    \caption{Positive energy level curves on the left and negative on the right, versus $m_1$ versus $\eta$. In (a-b) we consider $n=1$, $l=0$, and $\alpha=0.01$. In (c-d) we also use $n=1$, $l=0$, but set $\alpha=0.1$. Finally, in (e-f), we again set $\alpha=0.1$, for the case where $n=2$ and $l=2$. Source: The authors.}
    \label{fig:niveis_E_versus_m1_eta_3D}
\end{figure}

In Fig. \ref{fig:niveis_E_versus_m1_eta_3D}, we also present the energy level curves, with positive energies shown on the left and negative energies on the right, again as a function of $m_1$, but now with $\eta$ on the horizontal axis. In graphs (a-b), we consider the parameters $n=1$, $l=0$, and $\alpha=0.01$. In graphs (c-d), we maintain $n=1$ and $l=0$ while imposing $\alpha=0.1$. In graphs (e-f), we again adopt $\alpha=0.1$, considering the case where $n=2$ and $l=2$. We immediately notice that increasing $\eta$ (keeping $m_1$ fixed) makes the energies more positive, resulting in an increase in the magnitude of $E^+$ and a decrease in $E^-$, leading to $E^+$ and $E^-$ level curves that are mirrored to each other. The increase in $m_1$ (keeping $\eta$ fixed) results in a decrease, in magnitude, of both $E^+$ and $E^-$, such that both move away from the unit that represents $m_0c^2$. Furthermore, we can observe an increase in the magnitudes of the energies as $n$ and $l$ increase for each configuration point $(m_1,\eta)$. Moreover, it is evident that an increase in $\alpha$, at a given configuration point $(m_1,\eta)$, results in an increase in the magnitudes of energies $E^+$ and $E^-$.

\begin{figure}[!h]
    \centering
    \includegraphics[width=0.6\linewidth]{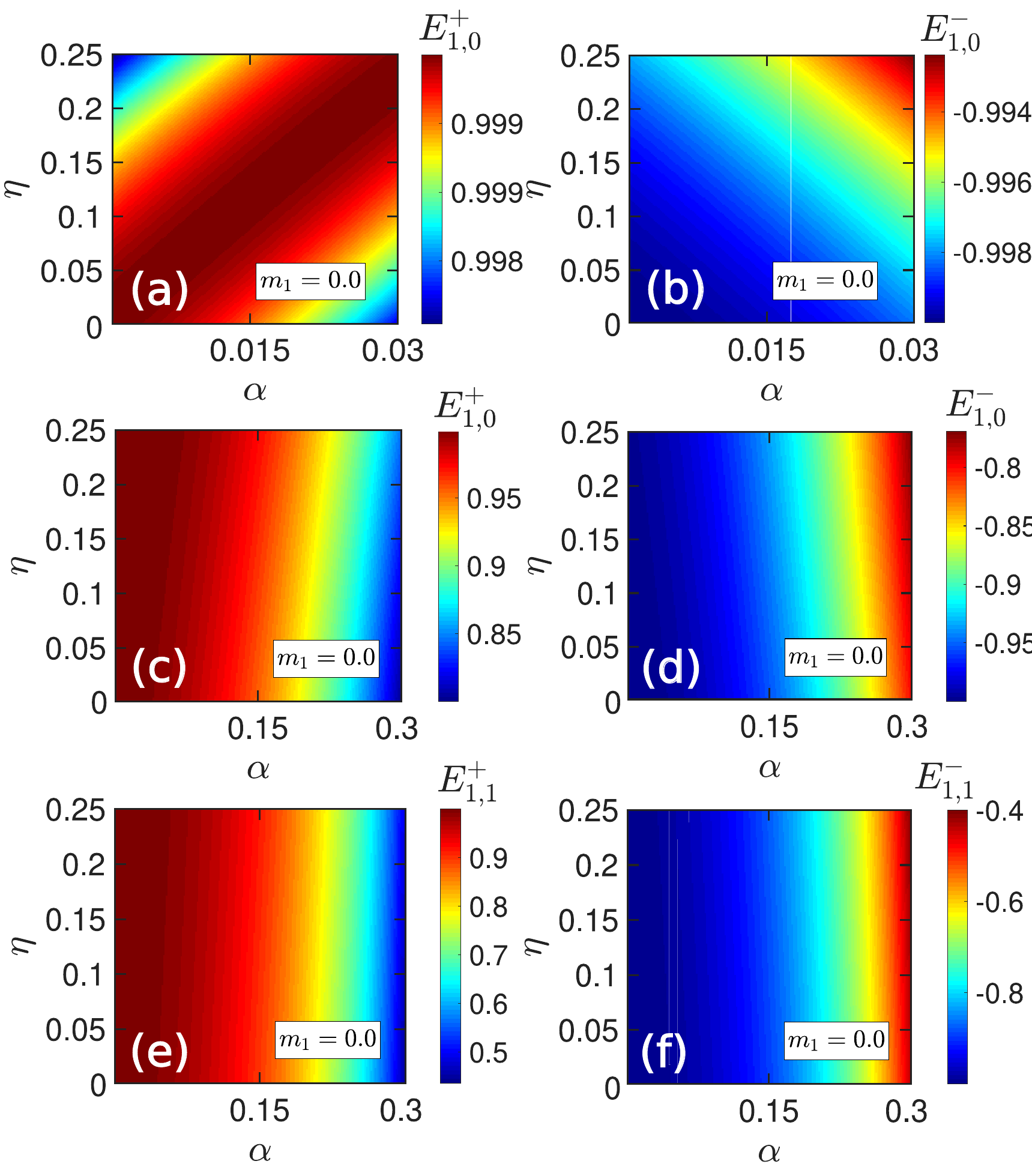}
    \caption{Positive energy level curves on the left and negative on the right, versus $\eta$ versus $\alpha$. In (a-d), we consider $n=1$, $l=0$, and $m_1=0.0$, with $\alpha$ varying in (a-b) over a smaller scale from $0$ to $0.03$, and in (c-d) over a larger scale from $0$ to $0.3$. In (e-f), we again set $m_1=0.0$, for the case where $n=1$ and $l=1$. Source: The authors.}
    \label{fig:niveis_E_versus_eta_alpha_m1_00_3D}
\end{figure}

In Fig. \ref{fig:niveis_E_versus_eta_alpha_m1_00_3D}, we present the energy level curves, with positive energies shown on the left and negative energies on the right, as a function of $\eta$ and $\alpha$. In graphs (a-d), we consider the parameters $n=1$, $l=0$, and $m_1=0.0$, with $\alpha$ varying in (a-b) a smaller scale from $0$ to $0.03$ and in (c-d) a larger scale from $0$ to $0.3$. In graphs (e-f), we maintain $m_1=0.0$ for the case where $n=1$ and $l=1$. We observe that in the small scale of $0\leq\alpha\leq0.03$ when we increase the value of $\alpha$ to $\eta\approx0.03$, the energies decrease in magnitude. Conversely, when increasing the value of $\alpha$ within the same scale but for values of $\eta\approx0.25$, we see that the magnitude of $E^-$ continues to decrease while the magnitude of $E^+$ begins to increase. Still within this range of $0\leq\alpha\leq0.03$, when we increase $\eta$ with $\alpha\approx0.01$ ($\alpha\approx0.03$), we find that $E^+$ decreases (increases) in magnitude, while $E^-$ increases in magnitude regardless of the value of $\alpha$. On the other hand, when we analyze a larger scale encompassing values of $0\leq\alpha\leq0.3$, we observe that regardless of the value of $\eta$, increasing $\alpha$ leads to a decrease in the magnitudes of the energies.

\begin{figure}[!h]
    \centering
    \includegraphics[width=0.6\linewidth]{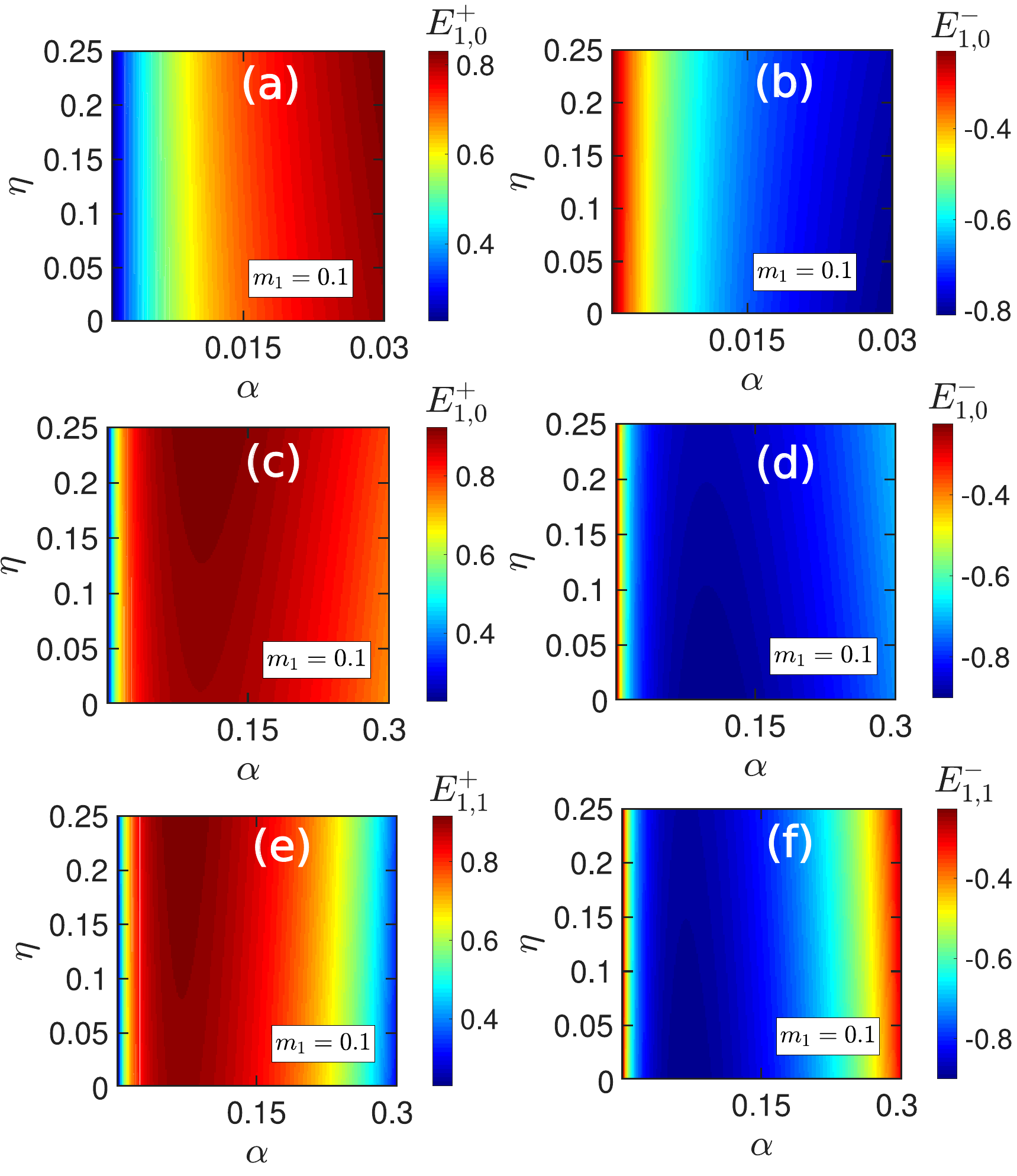}
    \caption{Positive energy level curves on the left and negative on the right, versus $\eta$ versus $\alpha$. In (a-d), we consider $n=1$, $l=0$, and $m_1=0.1$, with $\alpha$ varying in (a-b) over a smaller scale from $0$ to $0.03$, and in (c-d) over a larger scale from $0$ to $0.3$. In (e-f), we again set $m_1=0.1$, for the case where $n=1$ and $l=1$. Source: The authors.}
    \label{fig:niveis_E_versus_eta_alpha_m1_01_3D}
\end{figure}

In Fig. \ref{fig:niveis_E_versus_eta_alpha_m1_01_3D}, we again present the energy level curves, with positive energies shown on the left and negative energies on the right, as functions of $\eta$ and $\alpha$, now considering the effect of the PDM obtained with $m_1=0.1$. In graphs (a-d), we consider the parameters $n=1$, $l=0$, and $m_1=0.1$, with $\alpha$ varying in (a-b) a smaller scale from $0$ to $0.03$ and in (c-d) a larger scale from $0$ to $0.3$. In graphs (e-f), we maintain $m_1=0.1$ for the case where $n=1$ and $l=1$. We observe that the PDM eliminates the inversion behavior present in the scale of $0\leq\alpha\leq0.03$, such that the increase in $\alpha$ leads to an increase in the magnitudes of the energies, regardless of the value of $\eta$. Similarly, the PDM causes an increase in $\eta$ (keeping $\alpha$ fixed) to result in an increase in the magnitude of $E^+$ and a decrease in $E^-$, making both energies more positive. Conversely, when we assess a larger scale encompassing $0\leq\alpha\leq0.3$, we note that the increase in $\alpha$ (keeping $\eta$ constant) results in an increase in the magnitudes of the energies until a certain inversion point, beyond which increasing $\alpha$ leads to a decrease in the magnitudes of $E^+$ and $E^-$. Furthermore, the larger the numbers $n$ and $l$, the closer the inversion point is to $\alpha=0$.

\begin{figure}[!h]
    \centering
    \includegraphics[width=0.6\linewidth]{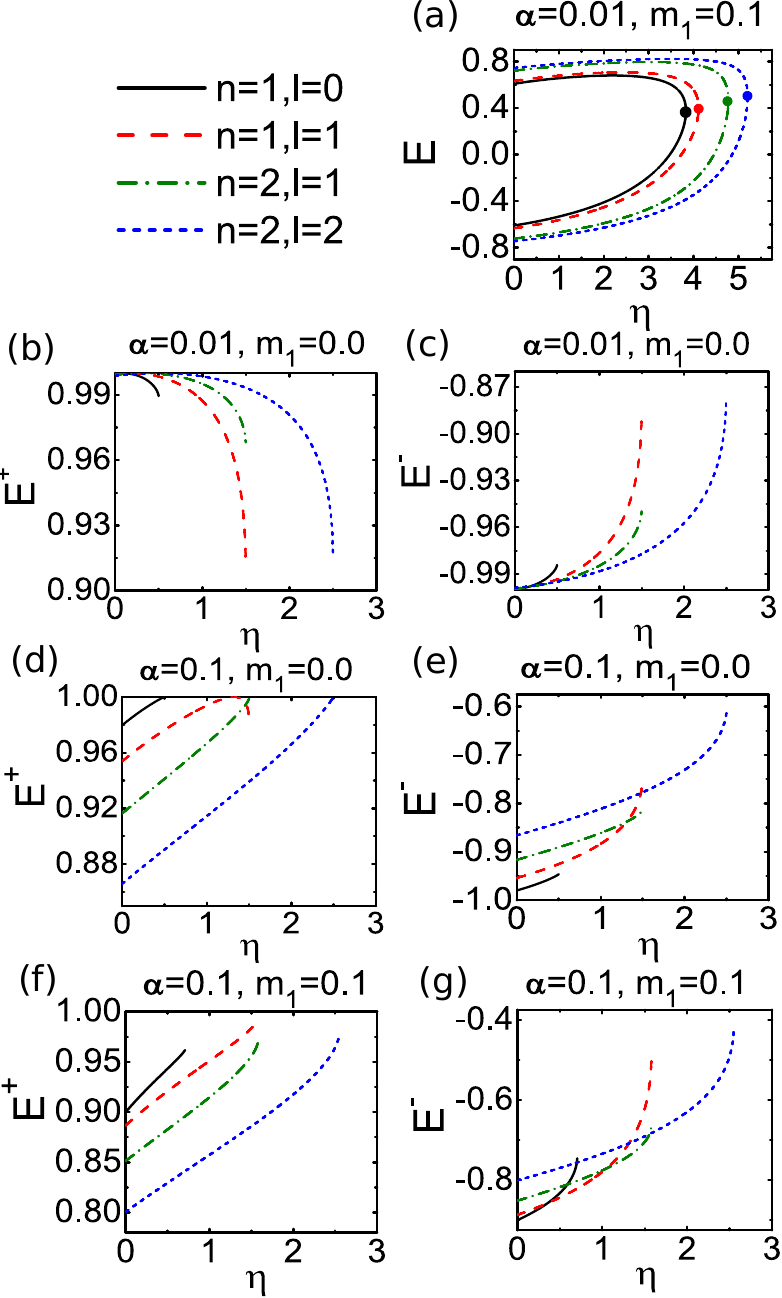}
    \caption{Energy levels versus $\eta$, for: $n=1$ and $l=0$ (solid black lines); $n=1$ and $l=1$ (red dashed lines); $n=2$ and $l=0$ (green dash-dotted lines); and $n=2$ and $l=2$ (blue dotted lines). In (a), we use $\alpha=0.01$ and $m_1=0.1$. The intersection points of the eigenvalue curves $E^+$ and $E^-$ are represented in the corresponding colors of each curve shown. In (b-c), we also use $\alpha=0.01$ but set $m_1=0.0$ for $E^+$ and $E^-$, respectively. In (d-e), we again use $m_1=0.0$ but set $\alpha=0.1$ for $E^+$ and $E^-$, in that order. In (f-e), we consider $\alpha=0.1$, but revert to setting $m_1=0.0$. Source: The authors.}
    \label{fig:graf_E_versus_eta}
\end{figure}

To better understand the evolution of energies for different configurations of $n$ and $l$, varying $\eta$ and $\alpha$, as well as considering whether $m_1 \neq 0$, we show in Fig. \ref{fig:graf_E_versus_eta} the energy levels as a function of $\eta$. The curves correspond to: $n=1$ and $l=0$ (solid black lines); $n=1$ and $l=1$ (dashed red lines); $n=2$ and $l=0$ (dot-dashed green lines); and $n=2$ and $l=2$ (dotted blue lines). In graph (a), we used $\alpha=0.01$ and $m_1=0.1$. The intersection points of the eigenvalue curves $E^+$ and $E^-$ are represented in the corresponding colors of each presented curve. In graphs (b-c), we used $\alpha=0.01$, but we considered $m_1=0.0$ for $E^+$ and $E^-$, respectively. In graphs (d-e), we maintained $m_1=0.0$ but considered $\alpha=0.1$ for $E^+$ and $E^-$, in the appropriate order. In graph (f), we again considered $\alpha=0.1$ but returned to $m_1=0.0$. Initially, we observe critical points of $\eta$ at which the energies $E^+$ and $E^-$ become imaginary, thus not represented in the spectrum. This behavior is recurrent in the level curves obtained by the Klein-Gordon formalism. With that, we will first analyze the cases with $\alpha=0.01$ (a-c), where we identify that the PDM highlights the symmetry breaking between the eigenvalues $E^+$ and $E^-$. This symmetry breaking exists for $m_1=0$ but is not as evident as in $m_1 \neq 0$. Additionally, $m_1 \neq 0$ also results in the closure of the energy gap for critical values of $\eta$, which is always present for $m_1=0$. When analyzing $E^+$ (b) and drawing vertical imaginary lines, we notice that the larger the values of $n$ and $l$, the greater the corresponding energy value. On the other hand, when we analyze the cases for $\alpha=0.1$ (d-f), we note behaviors significantly different from those observed for $\alpha=0.01$. This resembles the inversion behavior evidenced in the discussion of the level curves when the scale of $\alpha$ was on the order of the scale of $\eta$. Note that, for $\alpha=0.1$, as $\eta$ increases, the level curves of $E^+$ approach 1 ($m_0c^2$), instead of moving away from it, as occurred for $\alpha=0.01$. Furthermore, in this case, the PDM does not close the gap.

\begin{figure}
    \centering
    \includegraphics[width=0.6\linewidth]{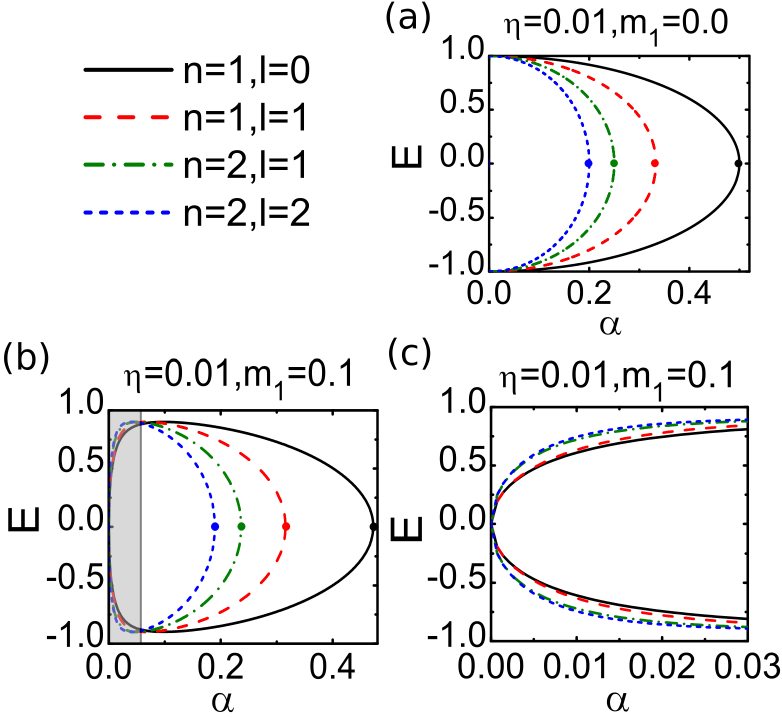}
    \caption{Energy levels versus $\alpha$, for: $n=1$ and $l=0$ (solid black lines); $n=1$ and $l=1$ (red dashed lines); $n=2$ and $l=0$ (green dash-dotted lines); and $n=2$ and $l=2$ (blue dotted lines). In (a), we use $\eta=0.01$ and $m_1=0.0$. In (b), we also use $\eta=0.01$ but set $m_1=0.1$. The intersection points of the eigenvalue curves $E^+$ and $E^-$ are represented in the corresponding colors of each curve shown. The shaded region is shown in more detail in (c). Source: The authors.}
    \label{fig:graf_E_versus_alpha}
\end{figure}

To analyze the behavior of the energy spectrum concerning $\alpha$, we present in Fig. \ref{fig:graf_E_versus_alpha} the energy levels as a function of $\alpha$ for: $n=1$ and $l=0$ (solid black lines); $n=1$ and $l=1$ (dashed red lines); $n=2$ and $l=0$ (dot-dashed green lines); and $n=2$ and $l=2$ (dotted blue lines). In graph (a), we used $\eta=0.01$ and $m_1=0.0$. In graph (b), we also used $\eta=0.01$, but we admitted $m_1=0.1$. The intersection points of the eigenvalue curves $E^+$ and $E^-$ are represented in the corresponding colors of each presented curve. The shaded region is shown in more detail in graph (c). Thus, we identify that when the mass does not depend on position, the energies start from 1 ($m_0c^2$) at $\alpha=0$, so that $E^+$ and $E^-$ decrease in magnitude until closing the energy gap at the critical points of $\alpha$, where both energies become imaginary. On the other hand, the PDM requires that at $\alpha=0$ the energies start from zero instead of 1 and that they increase in magnitude up to an inversion point ($\alpha \approx 0.03$, $E<1$), after which $E^+$ and $E^-$ decrease in magnitude until they close the energy gap at the critical point where they become imaginary.

\begin{figure}[!h]
    \centering
    \includegraphics[width=0.6\linewidth]{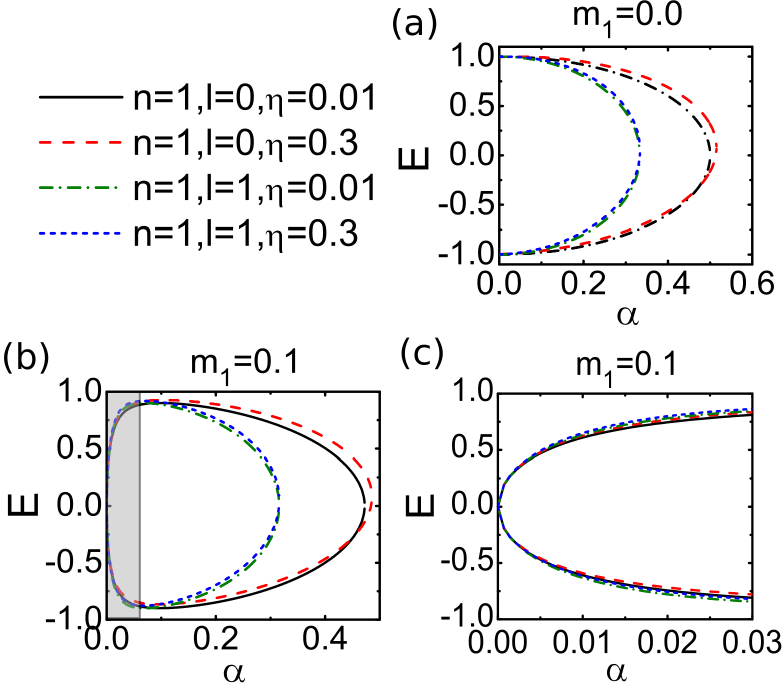}
    \caption{Energy levels versus $\alpha$, for: $n=1$, $l=0$, and $\eta$ (solid black lines); $n=1$, $l=0$, and $\eta=0.3$ (red dashed lines); $n=1$, $l=1$, and $\eta=0.01$ (green dash-dotted lines); and $n=1$, $l=1$, and $\eta=0.3$ (blue dotted lines). In (a), we use $m_1=0.0$. In (b), we set $m_1=0.1$, and the shaded region is shown in more detail in (c). Source: The authors.}
    \label{fig:graf_E_versus_alpha_comparado}
\end{figure}

To evaluate the effects on the energy spectrum behavior concerning $\alpha$, when we assume a larger value of $\eta=0.3$, we present in Fig. \ref{fig:graf_E_versus_alpha_comparado} the energy levels as a function of $\alpha$ for: $n=1$, $l=0$ and $\eta$ (solid black lines); $n=1$, $l=0$ and $\eta=0.3$ (dashed red lines); $n=1$, $l=1$ and $\eta=0.01$ (dot-dashed green lines); and $n=1$, $l=1$ and $\eta=0.3$ (dotted blue lines). In graph (a), we used $m_1=0.0$. In graph (b), we imposed $m_1=0.1$, and the shaded region is shown in more detail in graph (c). Thus, we can see that the effects of the PDM are maintained, but the effect of increasing $\alpha$ causes $E^+$ to increase as $n$ and $l$ increase, while $E^-$ decreases in magnitude with the increase of $n$ and $l$. This behavior does not shift the spectrum as a whole; it merely causes a deformation, as mentioned, which is also evidenced by the change in the critical point where the energy gap closes and the energies become imaginary.

\begin{figure}[!h]
    \centering
    \includegraphics[width=0.6\linewidth]{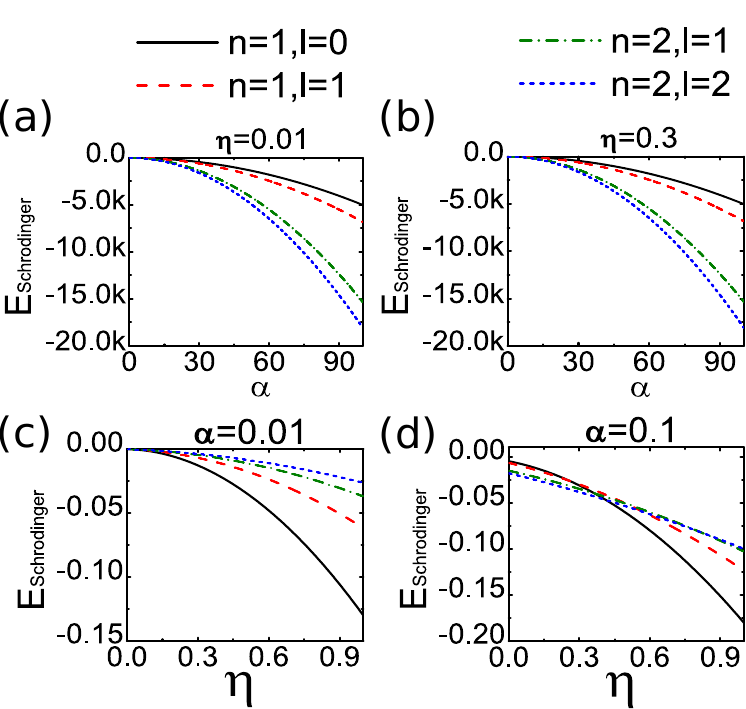}
    \caption{Energy levels according to the Schrödinger model for: $n=1$ and $l=0$ (solid black lines); $n=1$ and $l=1$ (red dashed lines); $n=2$ and $l=0$ (green dash-dotted lines); and $n=2$ and $l=2$ (blue dotted lines). In (a-b), we plot the energies versus $\alpha$, using $\eta=0.01$ and $\eta=0.0$, respectively. In (b), we also plot the energies versus $\eta$, setting $\alpha=0.01$ and $\alpha=0.1$, respectively. Source: The authors.}
    \label{fig:E_shrodinger}
\end{figure}

Finally, for comparison purposes, we show in Fig. \ref{fig:E_shrodinger} the energy levels obtained by the Schrödinger model for: $n=1$ and $l=0$ (solid black lines); $n=1$ and $l=1$ (dashed red lines); $n=2$ and $l=0$ (dot-dashed green lines); and $n=2$ and $l=2$ (dotted blue lines). In (a-b), we sketch the energies as a function of $\alpha$, adopting $\eta=0.01$ and $\eta=0.0$, respectively. In (b), we sketch the energies as a function of $\eta$, imposing $\alpha=0.01$ and $\alpha=0.1$, in that order. With this, we note that, unlike the Klein-Gordon formalism, we obtain real energy for any value of $\alpha$ and $\eta$, which we make evident by the large scale considered for $\alpha$. Furthermore, the energy levels start at $E=0$, increasing in magnitude, without any inversion point.

Curiously, some behaviors analogous to those obtained for Klein-Gordon stand out. In the energy spectrum as a function of $\alpha$, we find that the energy decreases (becomes more negative) as $n$ and $l$ increase, just like $E^+$ in Fig. \ref{fig:graf_E_versus_alpha}, which decreases from 1 $(m_0c^2)$ to $0$ for $m_1=0$. In the energy spectrum as a function of $\eta$, for $\alpha=0.01$, we find that the energy increases (in the positive sense of $E$) as $n$ and $l$ grow, which was also identified for the case of $\alpha=0.01$ and $m_1=0$ in Fig. \ref{fig:graf_E_versus_eta}. Even more curious is the case of $\alpha=0.1$ in Schrödinger, which presents the inverse behavior of energy when we vary $\eta$ and increase $n$ and $l$. The energy decreases (becomes more negative) as we take larger values of $n$ and $l$, but this behavior reverses from a certain value of $\eta$, for which the energy increases as $n$ and $l$ increase. This last behavior, where the energy is higher for larger $n$ and $l$, is also evidenced in the solution $E$ in Fig. \ref{fig:graf_E_versus_eta} for $\alpha=0.1$ and $m_0$. In the case of $\alpha=0.1$, we observe level crossings in both the Klein-Gordon and Schrödinger models, each assuming a position-independent mass.

\section{Conclusions}

In this article, we studied the effects of PDM on the bound state solutions of the Klein-Gordon particle subjected to the short-range Yukawa potential, starting from a mathematical expression already present in the literature, which has not been explored graphically or physically, as proposed in this work.

The study of eigenfunctions revealed that the presence of a mass function $m(r)$ renders the solutions (positive and negative) of a state with $n \leq 6$ indistinguishable from each other.

The graphical analysis of the eigenenergies, considering the variation of the terms governing (i) the PDM ($m_1$), (ii) the intensity ($\eta$), and (iii) the shielding ($\alpha$) of the Yukawa potential, showed that the energy spectra as a function of $\eta$ and $\alpha$ exhibit critical points at which positive and negative energies become imaginary, as characteristic of the solutions of the Klein-Gordon equation. Additionally, we observed that the behavior of the energies can be significantly different when $\eta$ and $\alpha$ are of the same order of magnitude or not. Curiously, we found that, in the energy spectrum as a function of $\eta$, PDM causes the closure of a gap at the critical point where $E^+$ and $E^-$ become imaginary. Moreover, in the energy spectrum as a function of $\alpha$, PDM requires that the energies be zero at $\alpha=0$, instead of being equal to $m_0c^2$, as is the case for the mass independent of position. Finally, we identified similarities between the solutions of the Klein-Gordon equation and the Schrödinger equation for the Yukawa potential and mass independent of position. We noted that both behave similarly regarding the evolution of the Schrödinger energy $E$ and the Klein-Gordon energy $E^+$ as the values of $(n, l)$ increase.




\end{document}